# BIRADS Features-Oriented Semi-supervised Deep Learning for Breast Ultrasound Computer-Aided Diagnosis

**Erlei Zhang[1,2], Stephen Seiler[3], Mingli Chen[2], Weiguo Lu[2*], Xuejun Gu[2*]**

[1]College of Information Science and Technology, Northwest University, Xi' an 710069, China
[2]Medical Artificial Intelligence and Automation Laboratory, Department of Radiation Oncology, University of Texas Southwestern Medical Center, Dallas, TX, 75390, USA
[3]Department of Radiology, University of Texas Southwestern Medical Center, Dallas, TX, 75390, USA

*E-mails: Weiguo.Lu@utsouthwestern.edu; Xuejun.Gu@utsouthwestern.edu

**Abstract**

We propose a novel BIRADS-SSDL network that integrates clinically-approved breast lesion characteristics (BIRADS features) into a task-oriented Semi-Supervised Deep Learning (SSDL) for accurate diagnosis on ultrasound (US) images with a small training dataset. Breast US images are converted to BIRADS-oriented Feature Maps (BFMs) using a distance-transformation coupled with a Gaussian filter. Then, the converted BFMs are used as the input of an SSDL network, which performs unsupervised Stacked Convolutional Auto-Encoder (SCAE) image reconstruction guided by lesion classification. This integrated multi-task learning allows SCAE to extract image features with the constraints from the lesion classification task, while the lesion classification is achieved by utilizing the SCAE encoder features with a convolutional network. We trained the BIRADS-SSDL network with an alternative learning strategy by balancing reconstruction error and classification label prediction error. To show the effectiveness of our approach, we evaluated it using two breast US image datasets. We compared the performance of the BIRADS-SSDL network with conventional SCAE and SSDL methods that use the original images as inputs, as well as with an SCAE that use BFMs as inputs. Experimental results on two breast US datasets show that BIRADS-SSDL ranked the best among the four networks, with classification accuracy around 94.23±3.33% and 84.38±3.11% on two datasets. In the case of experiments across two datasets collected from two different institution/and US devices, the developed BIRADS-SSDL is generalizable across the different US devices and institutions without overfitting to a single dataset and achieved satisfactory results. Furthermore, we investigate the performance of the proposed method by varying model training strategies, lesion boundary accuracy, and Gaussian filter parameter. Experimental results showed that pre-training strategy can help to speed up model convergence during training but no improvement of classification accuracy on testing dataset. Classification accuracy decreases as segmentation accuracy decreases. The proposed BIRADS-SSDL achieves the best results among the compared methods in each case and has the capacity to deal with multiple different datasets under one model. Compared with state-of-the-art methods, BIRADS-SSDL could be promising for effective breast US computer-aided diagnosis using small datasets.

# 1. Introduction

Breast cancer is the most common malignancy in women in the United States and the second leading cause of cancer death for women worldwide (Cheng *et al.*, 2010). Early detection and diagnosis are key to improving patient survival and quality of life, as they provide early and flexible treatment options (Osman and Yap, 2018). Breast ultrasound (US) is a widely adopted early breast cancer diagnosis imaging modality that has the advantages of being non-invasive, safe, efficient, and relatively inexpensive (Haynes and Moghaddam, 2010; Lee *et al.*, 2008). However, the main limitation of breast US is operator dependence(Hwang *et al.*, 2005). Many studies have applied Computer-Aided Diagnosis (CAD) to breast US to assist radiologists and improve diagnostic accuracy (Huang *et al.*, 2018).

In general, an US CAD system consists of four phases: image preprocessing, lesion segmentation, feature extraction, and classification. Image preprocessing is mainly designed for denoising and contrast enhancement. Segmentation focuses on separating the lesion region from the background and other tissue (Horsch *et al.*, 2001; Noble and Boukerroui, 2006). Feature extraction abstracts clinical characteristics from the segmented lesion, and classification utilizes the extracted features to differentiate the benign from the malignant. In traditional US CAD systems, most of the features are hand-crafted (Shen *et al.*, 2007; Flores *et al.*, 2015; Prabusankarlal *et al.*, 2015), considering radiomics, morphologic, and pathologic knowledge. Such feature extraction relies on clinical experience, and extracted features might not be robust in general. Over the years, clinical practice has accumulated knowledge regarding lesion characteristics for manual classification, which could be powerful for accurate diagnosis if incorporated into CAD. The Breast Imaging Reporting and Data System (BIRADS) was proposed by the American College of Radiology to help radiologists consistently describe and evaluate clinical lesions (D'orsi *et al.*, 1998). Descriptive terms of BIRADS features are clinically-approved lesion characteristics designed to quantify lesions by shape, contour attributes, internal echo patterns, and the architecture of the surrounding tissues.

In recent years, Deep Learning (DL) has been increasingly adopted in medical image analysis (Zou *et al*., 2019; Gessert *et al.*, 2019; Mishra *et al.*, 2018; Yap *et al.*, 2018a; Wang *et al.*, 2016; Shen *et al.*, 2017; Huynh *et al.*, 2016; Litjens *et al.*, 2017). DL-based methods have shown promising performance in computer vision tasks. Those methods not only support automatically representative and discriminative feature learning, but they also enable unsupervised feature learning (Bengio *et al.*, 2013). Deep Convolutional Neural Networks (CNN) have been used successfully for image segmentation (Ronneberger *et al.*, 2015) and classification (Hou *et al.*, 2016) tasks to achieve state-of-the-art performance. DL has also been applied in breast US diagnosis (Bian *et al.*, 2017; Shan *et al.*, 2016; Lei *et al.*, 2018; Shi *et al.*, 2016; Singh *et al.*, 2016; Zhang *et al.*, 2016). Cheng *et al*. (Cheng *et al.*, 2016) utilized unsupervised deep stacked Auto-Encoder (AE)-based methods to extract high-level features and supervised fine-tuning for breast US image classification. Han *et al*. (Han *et al.*, 2017) utilized the GoogLeNet pre-trained on gray natural images to classify breast US images with high accuracy. Antropova *et al*. (Antropova *et al.*, 2017) used ImageNet-pretrained CNNs to extract and pool low- to mid-level features and combine them with hand-crafted features to achieve accurate diagnoses on three imaging modality datasets. Most of those DL approaches require large amounts of data to train the models, whereas the pre-training strategy using nature images is designed for situations in which data are limited. However, the strategy of pre-training model on natural images faces challenges associated with the differences between the statistics of natural images and US images. Overall, DL has shown promising performance with automatic feature learning, but these approaches demand for large datasets and inefficiently utilize accumulated clinical knowledge.

In this paper, we report a novel BIRADS-SSDL network (architecture shown in Fig. 1) that incorporates the clinical knowledge of lesion characteristics (BIRADS features) and a task-oriented Semi-Supervised Deep Learning (SSDL) method to achieve accurate diagnosis on breast US images. The breast US images are converted to BIRADS-oriented Feature Maps (BFMs) using a distance-transformation coupled Gaussian filter. The BFMs not only keep the original US image information, but they also enhance the shape, lesion boundary, undulation, and angular characteristics of the lesion. Then, the BFMs are used as the input of an SSDL network, which performs a multi-task learning by integrating Stacked Convolutional Auto-Encoder (SCAE)-based unsupervised image feature extraction and diagnosis-oriented supervised lesion classification. This integrated multi-task learning allows SCAE to extract image features with the constraints from the lesion classification task, while the lesion classification is achieved by utilizing the SCAE encoder features with a convolutional network. The entire BIRADS-SSDL network is trained with an alternative learning strategy by balancing the reconstruction error and classification the label prediction error.

The paper is organized as follows: the proposed BIRADS-SSDL network is detailed in Section 2. Then, the proposed method's effectiveness is demonstrated by experimental results in Section 3. Finally, Section 4 provides a discussion and summary.

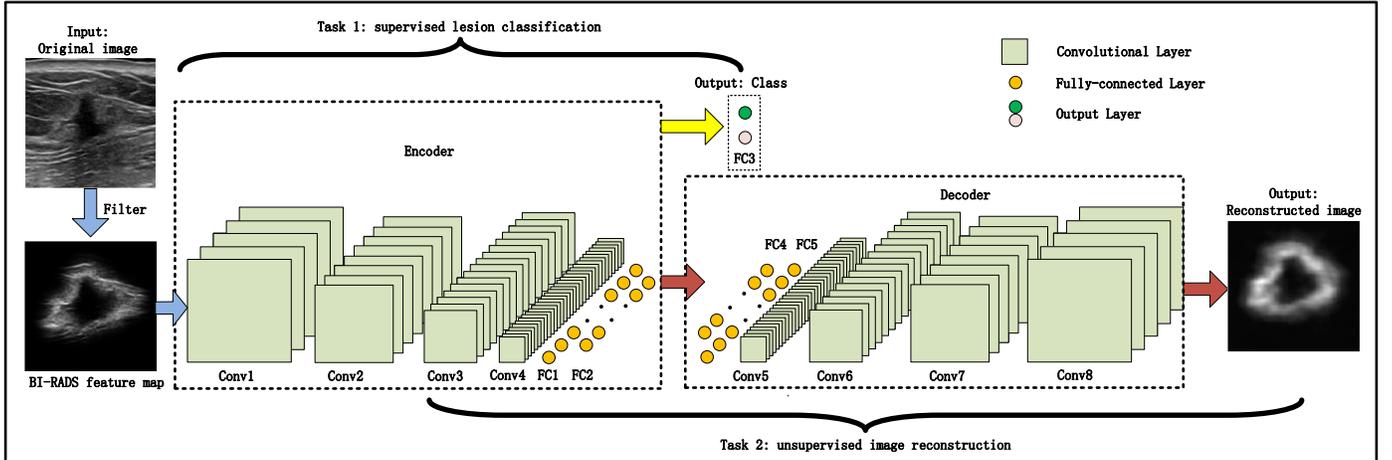

Figure. 1. Illustration of BIRADS-SSDL architecture, which consists of BIRADS feature map extraction and an SSDL network. The SSDL network integrates SCAE-based unsupervised image feature extraction with diagnosis-oriented supervised lesion classification.

## 2. Methods and Materials

### 2.1 BIRADS Features

BIRADS features consist of shape, orientation, margin, lesion boundary, echo pattern, and posterior acoustic feature classes (Shen *et al.*, 2007), which help radiologists grade the clinical findings, and evaluate their reliability against the pathological results. Fig. 2(a) shows a sample US image with the lesion boundary marked. The undulation and angular characteristics of the lesion, shown in Fig. 2(b), are two important BIRADS features for differentiating benign from malignant lesions. An undulation feature can be expressed as the number of significant lobulated areas partitioned between the lesion boundary and its maximum inscribed circle (blue circle). The angular characteristics can be detected through the local maxima (green dotted line) in each lobulated area on the distance map. In addition, the abrupt degree characteristic is usually calculated by the average gray intensities between the surrounding tissue and the lesion exterior, as shown in Fig. 2(c).

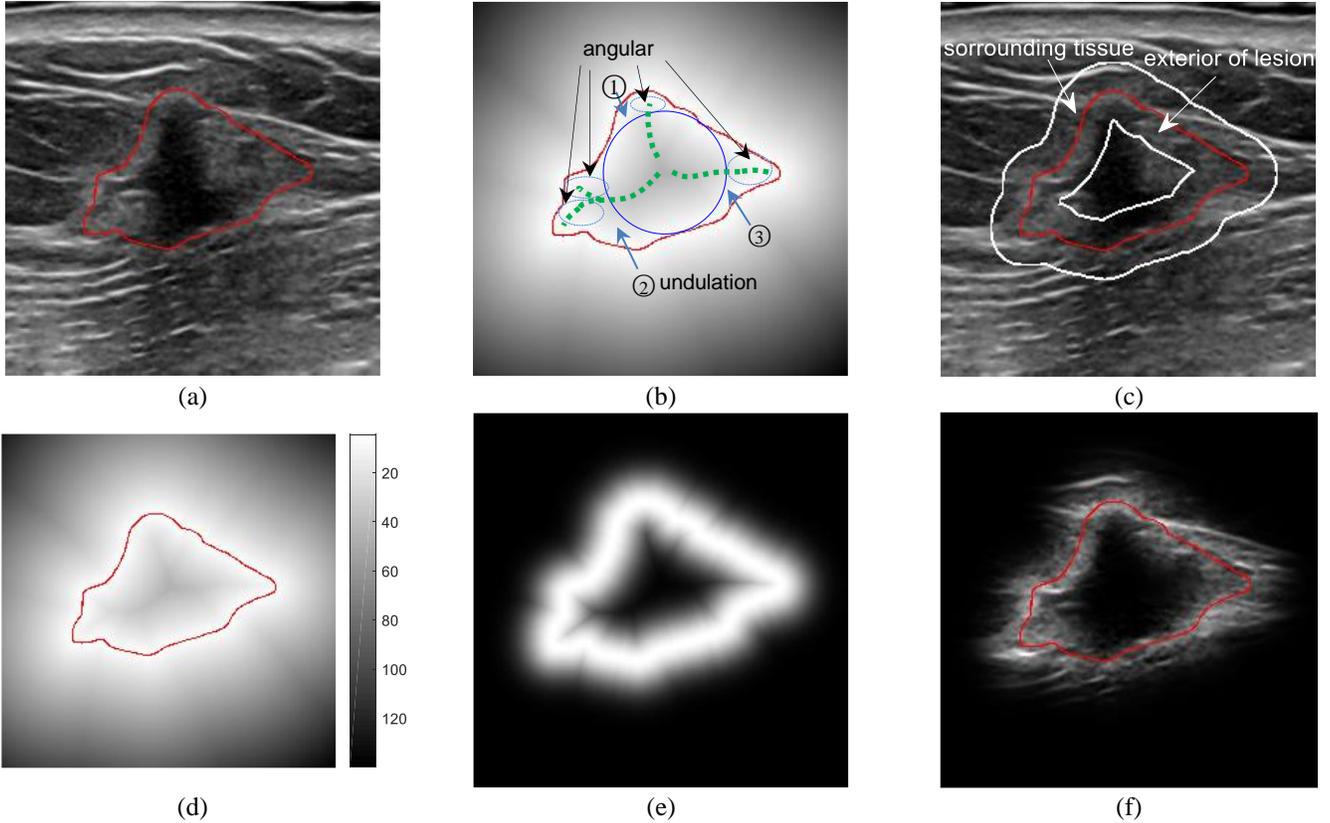

Figure. 2. An example of BIRADS-oriented feature map for capturing lesion characteristics: (a) A malignant lesion with boundary (red line); (b) Undulation and angular characteristics: the number of significant lobulated areas, and the number of the local maxima on the distance map; (c) Abrupt degree: the average gray intensities of the surrounding tissue and the lesion's exterior; (d) The distance map is represented by the grayscale, in which the lighter the gray, the smaller the distance to the boundary; (e) The outputs of **DTGF**: A distance-transformation coupled Gaussian filter with σ = 20; (f) BFMs: The BIRADS feature map with σ = 20.

## 2.2 BIRADS-oriented Feature Maps

The experimental results in the previous study (Shen *et al.*, 2007) showed that the most significant correlation exist between angular characteristic and pathological results, substantial correlations appears in features of irregular shape, undulation characteristic and degree of abrupt interface. The diagnostic value of posterior acoustic feature and orientation feature are relatively lower. Using this study as a reference, we preprocess US images with a distance transformed coupled Gaussian filter, **DTGF**(·), which enhances shape and margin features (angular, undulation, and degree of abrupt interface), while keeps orientation, echo pattern and posterior features. **DTGF**(·) is defined as

$$\boldsymbol{DTGF}(\boldsymbol{p}) = e^{-\frac{Dist(p)^2}{\sigma^2}} \tag{1}$$

where the distance transform $\boldsymbol{Dist}(\boldsymbol{p}) = min\{ED(\boldsymbol{p}, \boldsymbol{q_i})\}$ represents the Euclidean distance between the image pixel $p$ and boundary pixel $q_i$. An example of a distance map represented by grayscale is shown in Fig. 2(d). σ is used to control the width of the region of surrounding tissue and the exterior lesion across the boundary. The $DTGF$ assigns different weights based on the distance of the pixel to the boundary and promotes attention to key areas. An example of $DTGF$ with σ = 20 is shown in Fig. 2(e).

With $\boldsymbol{DTGF}$, original breast US images $\boldsymbol{I}$ are converted to BFMs as follow:

$$BFMs = \boldsymbol{I} \cdot \mathrm{e}^{-\frac{Dist(p)^2}{\sigma^2}} \tag{2}$$

An example of BFMs is shown in Fig. 2(f). From the figure, it can be seen that the shape and boundary of the lesion are emphasized, and the undulation and angular characteristics based on the distance map are well reflected in the BFMs. The converted BFMs not only keep the key information from the original US image but also enhance the lesion's shape, boundary, undulation, and angular characteristics. With the guidance of the BFMs, the advanced deep feature learning can focus on the clinically-assigned breast lesion characteristics.

*2.3 SCAE Neural Network*

A SCAE neural network follows an unsupervised encoder-decoder learning paradigm (Cheng *et al.*, 2016; Gondara, 2016). A standard AE network is a three-layer network that attempts to output an approximation of the input with an encoder and a decoder. The encoder maps the input $\{\boldsymbol{x}^{(n)} \in \mathfrak{R}^d\}_{n=1}^N$ to a hidden feature representation vector $\boldsymbol{h} \in \mathfrak{R}^{d'}$ through a nonlinear projection function (activation function) $\boldsymbol{f}_{en}(\cdot)$:

$$\boldsymbol{h} = \boldsymbol{f}_{en}(\boldsymbol{W}_{en} \cdot \boldsymbol{x} + \boldsymbol{b}_{en}) \tag{3}$$

Then, the decoder maps the hidden feature representation $\boldsymbol{h}$ to an output vector $\widehat{\boldsymbol{x}} \in \mathfrak{R}^d$:

$$\widehat{\boldsymbol{x}} = \boldsymbol{f}_{de}(\boldsymbol{W}_{de} \cdot \boldsymbol{h} + \boldsymbol{b}_{de}) \tag{4}$$

where $\widehat{\boldsymbol{x}}$ is expected to be an approximate reconstruction of the input $\boldsymbol{x}$. The model is learned to minimize reconstruction error:

$$\min_{\boldsymbol{\theta}_r} J(\boldsymbol{\theta}_r) = \frac{1}{N}\sum_{n=1}^N loss_r(\boldsymbol{x}^{(n)}, \widehat{\boldsymbol{x}}^{(n)}) + \gamma \cdot \boldsymbol{R}(\boldsymbol{\theta}_r) \tag{5}$$

where the reconstructing loss function $loss_r(\cdot)$ uses Euclidean distance $loss_r(\boldsymbol{x}, \widehat{\boldsymbol{x}}) = \|\boldsymbol{x} - \widehat{\boldsymbol{x}}\|_2^2$ and parameter regular term $\boldsymbol{R}(\boldsymbol{\theta}_r) = \|\boldsymbol{W}_{en}\|_F^2 + \|\boldsymbol{W}_{de}\|_F^2$.

Convolutional AE (CAE) (Masci *et al.*, 2011) is an extension of the standard AE that introduces a convolution operation between hierarchical connections. Unlike standard AE, the inputs of CAE are not restricted to one-dimensional vectors but can also be 2D images. CAE captures structural information and preserves the local spatiality of an image by sharing weights among all input locations. Similar to CNN, the hidden feature map $\boldsymbol{h}$ is given by

$$\boldsymbol{h} = \boldsymbol{f}_{en}(\boldsymbol{W}_{en} * \boldsymbol{x} + \boldsymbol{b}_{en}) \tag{6}$$

With the feature maps, the reconstruction of input is obtained using

6                                                                                                                                       E. Zhang *et al.*$$\hat{x} = f_{de}(W_{de} * h + b_{de}) \tag{7}$$

where * denotes the 2D convolution.

A SCAE network is formed by stacking several CAEs hierarchically. The input of the *i*+1-th layer is the feature map of the *i*-th layer:

$$\begin{cases} h^l = f_{en}^l(W_{en}^l * h^{l-1} + b_{en}^l) \\ h^1 = f_{en}^1(W_{en}^1 * x + b_{en}^1) \end{cases} \tag{8}$$

where $l = 2, \cdots, L$. The whole network is unsupervised trained in a greedy, layer-wised fashion by minimizing the reconstruction error of its input. Once all layers have been trained, a minimized reconstruction error means that the feature maps from the output of the encoder contain the most important information from the input. The traditional SCAE-based classifier includes two detached stages: 1. unsupervised learning for image reconstruction; 2. fine-tuning the classification network only with supervised learning. After stage 1 is finished, the classification stage removes the decoder and only preserves the input and the encoder. The output of the encoder is then fed into a softmax classifier or other classifiers with supervised training. The classic methods (Masci *et al.*, 2011) learn the softmax classifier or fine-tune the parameters of all the layers together with the labeled samples as inputs.

*2.4 BIRADS-SSDL Neural Network*

As mentioned in Section 2.C., representative features from a standard SCAE are learned mainly for image reconstruction. The final stage fine-tuned the parameters with supervised learning may have a limited impact on the classification task. More importantly, it is difficult to learn an effective CNN model directly with a small number of labeled samples.

Inspired by the multi-task network (Ghifary *et al.*, 2016), we developed a novel BIRADS-SSDL network, as shown in Fig. 1. Let $\{x^{(n)}, y^{(n)}\}_{n=1}^N$ be the labeled samples. $y \in \Re^K$ is a one-hot label vector. The lesion classification task is implemented by an encoder network and a classifier, as shown in Fig. 1, which can be expressed as

$$\begin{cases} h = f_{en}(W_{en} * x + b_{en}) \\ \hat{y} = f_c(h, \beta) \end{cases} \tag{9}$$

where $h$ is the output of the encoder, $f_c(\cdot)$ is a softmax classifier, and $\beta$ is a vector of parameters of the classifier to be learned. $\hat{y} \in \Re^K$ is the output of the classifier and ranges within [0, 1]. The objective function of classification is

$$\min_{\theta_c} J(\theta_c) = \frac{1}{N} \sum_{n=1}^{N} loss_c(y^{(n)}, \hat{y}^{(n)}) + \gamma \cdot R(\theta_c) \tag{10}$$

where $\theta_c = [W_{en}, b_{en}; \beta]$ is learning or tuning by the training dataset. Commonly, the loss function $loss_c(\cdot)$ uses the binary cross-entropy:

$$loss_c(y, \hat{y}) = -\sum_{k=1}^{K} \delta(y(k) = 1) \cdot \log(\hat{y}(k)) \tag{11}$$

where δ(·) is an indicative function (the value is 1 if $y(k)$ is equal to 1), and $K$ is the number of classes. Here, $K = 2$ because the lesion is either benign or malignant.

The image reconstruction pipeline is similar to the standard SCAE structure and consists of an encoder and a decoder, as shown in Fig. 1. Combined with the reconstruction pipeline, the objective function of the BIRADS-SSDL network is as follows:

$$\min_{\boldsymbol{\theta}} J(\boldsymbol{\theta}) = \frac{1}{N}\sum_{n=1}^{N} \lambda \cdot loss_c(\boldsymbol{y}^{(n)}, \widehat{\boldsymbol{y}}^{(n)}) + (1-\lambda) \cdot loss_r(\boldsymbol{x}^{(n)}, \widehat{\boldsymbol{x}}^{(n)}) + \gamma \cdot \boldsymbol{R}(\boldsymbol{\theta}) \qquad (12)$$

where $\boldsymbol{\theta} = [\boldsymbol{W}_{en}, \boldsymbol{b}_{en}; \boldsymbol{W}_{de}, \boldsymbol{b}_{de}; \boldsymbol{\beta}]$ and λ ∈ [0,1] is used to balance the classification and reconstruction tasks. The objective function is a convex optimization problem and can be achieved by alternative learning (Ghifary *et al.*, 2016). During feature learning, the encoding parameters are shared among both tasks, while the decoding parameters only participate in the reconstruction task.

In this paper, the classification pipeline has four convolutional layers: 8(Conv1), 16(Conv2), 32(Conv3), and 64(Conv4) @3×3 filters respectively, four max-pooling layers of size 2×2 after each convolutional layer, and three fully-connected layers (FC1, FC2, FC3), as shown in Fig. 1. The number of neurons in FC1 and FC2 is 256 and 64, respectively. The output layer FC3 has a softmax activation function with two neurons. The dropout (with a probability of 0.5) is applied after FC1 and FC2 to prevent overfitting. ReLU activations are used in all hidden layers. The reconstruction pipeline has an encoder and a decoder. The encoder is shared with the classification pipeline, including the four convolutional layers (Conv1, Conv2, Conv3, and Conv4) and two fully-connected layers (FC1 and FC2). The decoder has the inverse configuration of the encoder, including two fully-connected layers (FC4 and FC5), four pairs of convolution and upsampling layers, and a convolutional output layer with linear activations. The classification and reconstruction tasks are alternately updated via Adam with a learning rate of $3\times10^{-4}$ and the parameter λ equal to 0.5, stopping to update the network when the average reconstruction loss remains stable.

*2.5 Experimental Setups*

*2.5.1    Datasets*

Dataset I — Public UDIAT: We used a public breast B-Mode US image dataset, named UDIAT (Yap *et al.*, 2018b), to investigate the effect of the proposed methods. This dataset was collected in the UDIAT Diagnostic Centre of the Parc Taulí Corporation, Sabadell, Spain with a Siemens ACUSON Sequoia C512 system 17L5 HD linear array transducer (8.5 MHz). The average size of the images is 760×570 pixels, with a nominal pixel size of 0.084mm. The lesions were delineated by an experienced radiologist. In this study, 128 images with lesion sizes smaller than 512×512 pixels were selected and cropped to 512×512 centered on the lesions. These 128 images include 45 images with malignant lesions and 83 with benign lesions.

Dataset II — In-house clinical dataset: The in-house clinical dataset, named UTSW dataset, is a B-mode US breast image dataset collected at the University of Texas Southwestern Medical Center with a Philips iU22 scanner (Philips Medical Systems, equipped with a 12-5 MHz linear array transducer). The average size of the images is $870 \times 660$ pixels, with pixel size varying from 0.04 to 0.1mm (average pixel size is 0.068mm). The lesions were identified as benign or malignant based on the pathologic examination of a subsequent biopsy. Lesions on the images were marked with two or four boundary points. In this study, we selected 258 images from 144 patients, including 178 benign lesions and 80 malignant lesions. The images were resampled to a resolution of 0.084mm and cropped to 512×512 centered on the lesions. A marker-controlled watershed segmentation method was used to create the tumor boundary (Gomez *et al.*, 2010).

*2.5.2  Experimental Setups*

We designed three scenarios to evaluate BIRADS-SSDL's performance within and across the two datasets: 1) within the UDIAT dataset, 80% of the samples were randomly selected from benign and malignant lesions to form the training set, and the remaining 20% were used as the testing set; 2) within the UTSW dataset, 80% per class of samples were randomly chosen to construct a training set, and 20% of the samples were randomly chosen as the test set; 3) a combined training set was constructed from 80% of the samples from UDIAT and 80% of samples from UTSW dataset, and the remaining samples from each dataset were used as two testing sets;

We also characterize BIRAD-SSDL's performance on with 1) Pre-training strategy: 80% of the samples selected from the UTSW (UDIAT) dataset were used to pre-train the models, and the whole network was fine-tuned with 80% of the samples from UDIAT (UTSW), and then the models were tested on the remaining UDIAT (UTSW) images; 2) varied lesion segmentation accuracy and 3) various Gaussian filter parameter.

In all designed experiments, the gray values of pixels were normalized to [0, 1]. All algorithms were executed using Python in the environment of NVIDIA GeForce RTX 2080 Ti GPU and 64 GB of RAM.

*2.5.3  Performance Evaluation Metric*

In this paper, ACC, AUC, SEN, SPE, PPV, NPV, and MCC represent seven performance metrics (Powers, 2011; Zhou *et al.*, 2016): accuracy, area the under receiver operating characteristic curve, sensitivity, specificity, positive predictive value, negative predictive value, and Matthews correlation coefficient, respectively. In the experiments, TP is the number of true positives (malignant breast tumor), FN is the number of false negatives (benign breast tumor), TN is the number of true negatives, and FP is the number of false positives.

$$
\begin{aligned}
\text{SEN} &= \text{TP}/(\text{TP} + \text{FN}) \\
\text{SPE} &= \text{TN}/(\text{FP} + \text{TN}) \\
\text{PPV} &= \text{TP}/(\text{TP} + \text{FP}) \\
\text{NPV} &= \text{TN}/(\text{TN} + \text{FN}) \\
\text{ACC} &= (\text{TP} + \text{TN})/(\text{TP} + \text{FP} + \text{FN} + \text{TN}) \\
\text{AUC} &= 0.5 \cdot (\text{TP}/(\text{TP} + \text{FN}) + \text{TN}/(\text{TN} + \text{FP})) \\
\text{MCC} &= \frac{\text{TP} \cdot \text{TN} - \text{FP} \cdot \text{FN}}{\sqrt{(\text{TP} + \text{FN})(\text{TP} + \text{FP})(\text{TN} + \text{FN})(\text{TN} + \text{FP})}}
\end{aligned}
\tag{13}
$$

ACC measures the ratio of the number of samples correctly classified to the total number of samples. AUC indicates the trade-off between SEN and SPE, whose advantages are the robust description of the classifier's predictive ability. MCC gives a better evaluation than ACC when the numbers of negative samples and positive samples are unequal. The larger the value is, the better the performance of the classifier.

## 3  Results

### *3.1 Classification Results on Single Dataset*

To demonstrate the effectiveness of BIRADS-SSDL, we chose three SCAE-based methods for comparisons: 1) ORI-SCAE, which uses original images as inputs and a standard SCAE network with unsupervised learning for reconstruction, then adds three fully-connected layers (FC1, FC2, FC3) for diagnosis prediction. Fine-tuning with the labeled samples is performed on the whole network; 2) BIRADS-SCAE, which is similar to ORI-SCAE, but it uses with BFMs as the network inputs; and 3) ORI-SSDL, a semi-supervised learning method, like BIRADS-SSDL, that uses original images, not BFMs, as inputs.

Table 1 shows the classification results (mean ± standard deviation %) for the four methods on the UDIAT dataset. First, comparing the methods with different inputs (ORI-SCAE and BIRADS-SCAE, ORI-SSDL and BIRADS-SSDL), we found that BIRADS-based methods outperformed ORI-based methods from the perspective of the seven metrics. The BIRADS-based methods achieved ACC, AUC, and MCC values about 5%, 5%, and 9% higher, respectively, than the ORI-based methods, which indicates the advantage of using BIRADS-oriented feature maps. Second, comparing ORI-SSDL to ORI- SCAE and BIRADS-SSDL to BIRADS-SCAE, we found that the SSDL-based methods (ORI-SSDL and BIRADS-SSDL) obtained better results than the methods with the fine-tuning strategy (ORI-SCAE and BIRADS-SCAE), which means they learn more effective features for classification using unsupervised image reconstruction with the constraints from the lesion classification task. Overall, BIRADS-SSDL produces the best diagnosis results by taking advantage of BIRADS features and SSDL with small datasets.

Table 1. Classification results (mean ± std %) for different methods when training and testing on UDIAT.

| Method | ORI-SCAE | ORI-SSDL | BIRADS-SCAE | BIRADS-SSDL |
|---|---|---|---|---|
| ACC | 87.50±3.72 | 89.01±6.31 | 91.03±2.87 | **94.23±3.33** |
| AUC | 82.87±6.02 | 86.43±6.47 | 87.87±5.09 | **90.80±6.01** |
| MCC | 69.98±8.95 | 76.78±11.96 | 78.65±7.34 | **85.80±8.45** |
| SEN | 70.42±14.03 | 77.74±11.87 | 79.26±12.3 | **82.99±12.28** |
| SPE | 95.33±4.16 | 95.12±7.32 | 96.48±4.07 | **98.61±2.41** |
| PPV | 86.96±11.37 | 93.33±8.46 | 90.56±9.51 | **96.88±5.41** |
| NPV | 88.13±6.63 | 87.84±7.24 | 91.46±5.73 | **93.71±4.06** |

Table 2. Classification results (mean ± std %) for different methods when training and testing on UTSW dataset.

| Method | ORI-SCAE | ORI-SSDL | BIRADS-SCAE | BIRADS-SSDL |
|---|---|---|---|---|
| ACC | 80.77±4.03 | 81.09±1.73 | 82.21±2.59 | **84.38±3.11** |
| AUC | 72.38±4.79 | 71.78±4.02 | 71.55±2.24 | **77.14±6.61** |
| MCC | 52.57±7.01 | 48.95±7.22 | 53.10±7.21 | **60.01±8.96** |
| SEN | 50.94±12.61 | 51.09±8.89 | 47.15±13.65 | **60.83±15.37** |
| SPE | 93.45±5.16 | 92.47±2.86 | **95.94±2.35** | 93.82±3.99 |
| PPV | 81.54±9.95 | 72.11±8.82 | **83.65±9.75** | 81.71±8.87 |
| NPV | 81.41±5.06 | 83.19±2.67 | 82.15±3.35 | **85.95±3.57** |

Similar conclusions can be drawn from the classification results on the UTSW dataset, shown in Table 2. BIRADS-SSDL outperformed the other three compared methods in terms of MCC, ACC, and AUC. Unlike the UDIAT dataset, BIRADS-SSDL only outperformed the other methods in terms of ACC by about 2%~4%. While BIRADS-SSDL achieved a much higher SEN than the other methods. Moreover, BIRADS-SSDL outperformed the other methods in terms of MCC and AUC by about 5%. This indicates that the proposed BIRADS-SSDL achieves a better balance on the predictions of negative and positive samples than the other methods.

*3.2 Model Validation across Dataset*

Table 3 and 4 summarize the classification results across datasets with the same comparison as described in Section 3.A. Each method was trained on the combined UDIAT (randomly selected 80% of the samples) and UTSW datasets (randomly selected 80% of the samples), then tested on the remaining samples from each dataset, respectively. It can be seen that all the methods produced results similar to those shown in Table 1 and 2. Although two datasets collected by different manufacturers' devices have different characteristics, the proposed BIRADS-SSDL performed the best in all the comparisons. This indicates that, among the methods compared, BIRADS-SSDL method is more generalizable across different datasets without overfitting to single institution data.

Table 3. Classification results (mean ± std %) for different methods when training on a combined UDIAT and UTSW dataset and testing on UDIAT.

| Method | ORI-SCAE | ORI-SSDL | BIRADS-SCAE | BIRADS-SSDL |
|---|---|---|---|---|
| ACC | 85.58±2.67 | 88.66±3.22 | 90.77±1.88 | **92.95±3.45** |
| AUC | 82.53±5.52 | 84.29±4.12 | 85.73±3.48 | **88.39±5.02** |
| MCC | 67.13±7.03 | 73.99±6.79 | 73.51±6.67 | **81.41±7.66** |
| SEN | 74.01±12.81 | 71.88±9.16 | 75.43±7.62 | **79.56±10.68** |
| SPE | 91.06±4.22 | 96.70±3.75 | 96.03±2.02 | **97.22±2.78** |
| PPV | 81.88±9.25 | 92.28±8.29 | 83.81±10.58 | **93.39±6.70** |
| NPV | 87.57±5.86 | 87.88±4.22 | 91.92±4.35 | **93.30±3.97** |

Table 4. Classification results (mean ± std %) for different methods when training on a combined UDIAT and UTSW dataset and testing on UTSW.

| Method | ORI-SCAE | ORI-SSDL | BIRADS-SCAE | BIRADS-SSDL |
|---|---|---|---|---|
| ACC | 79.81±3.67 | 81.15±3.31 | 81.54±2.61 | **83.27±4.79** |
| AUC | 72.80±5.55 | 71.60±4.35 | 73.94±4.68 | **76.59±5.22** |
| MCC | 48.38±7.43 | 51.87±9.08 | 56.24±7.60 | **59.21±10.65** |
| SEN | 57.22±15.10 | 48.03±9.55 | 51.88±9.49 | **60.10±11.31** |
| SPE | 88.38±4.28 | 95.17±4.07 | **96.01±2.28** | 93.09±5.84 |
| PPV | 67.62±6.85 | 81.45±14.23 | **86.01±5.27** | 82.09±10.02 |
| NPV | 84.16±3.07 | 81.39±4.64 | 80.35±3.12 | **84.20±4.22** |

*3.3 Effect of model pre-training*

Further, all models were pre-trained on the UTSW (UDIAT) dataset, and the whole network was fine-tuned with 80% of the samples from UDIAT (UTSW), and then the models were tested on the remaining UDIAT (UTSW). The classification results are shown in Table 5, respectively. Compared with the results in Table 1 and 2, there is no obvious difference in the evaluation metrics. We observed the changes of $loss_r$ and $loss_c$ function during training the models on UDIAT and UTSW datasets, as shown in Fig. 3. It can be seen that the loss values of image reconstruction decrease rapidly with the iteration until it is relatively stable. The reconstruction losses of BIRADS-SSDL with the pre-trained strategy (denoted as transfer BIRADS-SSDL) are smaller than the losses of BIRADS-SSDL, and the convergence speed is faster. The losses of image classification show a similar trend. A pre-trained model can help BIRADS-SSDL speed up convergence and achieve smaller losses of image reconstruction.

Table 5. Classification results (mean ± std %) of BIRADS-SSDL and transfer BIRADS-SSDL

| Method | UDIAT | | UTSW | |
| --- | --- | --- | --- | --- |
| | BIRADS-SSDL | Transfer BIRADS-SSDL | BIRADS-SSDL | Transfer BIRADS-SSDL |
| ACC | **94.23±3.33** | 93.59±2.87 | **84.38±3.11** | 84.23±1.44 |
| AUC | 90.80±6.01 | **92.99±3.05** | 77.14±6.61 | **77.63±1.64** |
| MCC | 85.80±8.45 | **85.98±6.79** | 60.01±8.96 | **62.31±5.22** |
| SEN | 82.99±12.28 | **89.68±5.02** | **60.83±15.37** | 59.38±5.91 |
| SPE | **98.61±2.41** | 96.29±4.15 | 93.82±3.99 | **95.87±4.61** |
| PPV | **96.88±5.41** | 92.25±8.54 | 81.71±8.87 | **86.77±14.58** |
| NPV | 93.71±4.06 | **93.77±4.20** | **85.95±3.57** | 83.98±4.17 |

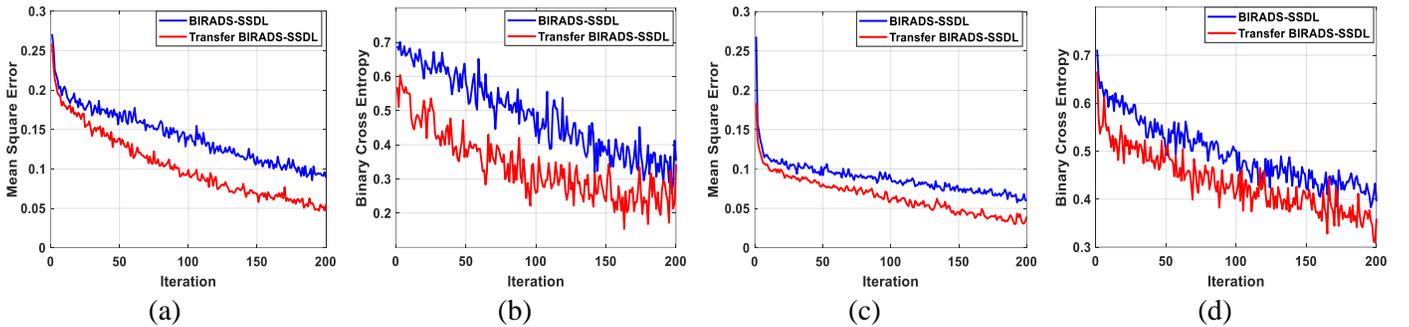

Figure 3. The loss values of BIRADS-SSDL and transfer BIRADS-SSDL during training on UDIAT and UTSW respectively: (a) and (b) are the loss values of image reconstruction and classification on UDIAT dataset; (c) and (d) are the loss values of image reconstruction and classification on UTSW dataset.

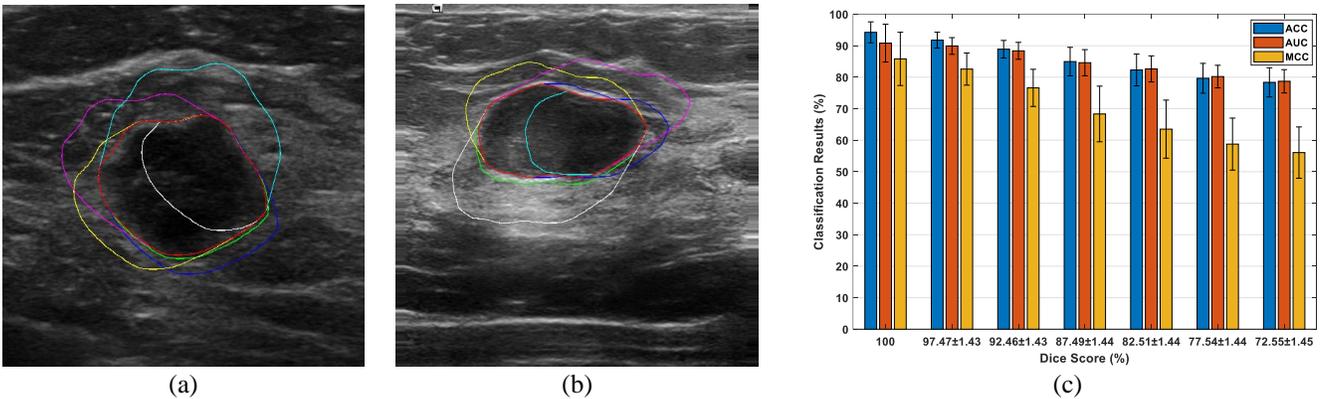

Figure 4. The effects of lesion contour accuracy on the performance of BIRADS-SSDL method. (a) and (b) are two lesion samples with tumor boundary: real-boundary is red, and fake-boundaries with different dice score (about 70% ~ 95%) are in different colors. (c) is classification results of BIRADS-SSDL corresponding to different lesion contour accuracy.

### 3.4 Effects of Lesion Contour Accuracy

In this paper, BFMs are extracted based on tumor boundary. Commonly, the more precise the boundary detection the better feature representation. In order to investigate the effects of lesion segmentation accuracy on the BIRADS-SSDL, we used morphological operator (dilate and erode image)(Gonzalez *et al.*, 2009) to modify real lesion boundary producing fake-boundary on the testing samples of UDIAT dataset. In our experiments, we use various structuring elements to produced 91873 testing images with fake-boundary. Fig. 4(a) and (b) shows the boundary of two lesions, including the dataset provided real-boundary which was delineated by radiologists and the fake-boundary. Then, we extracted BFMs based on fake-boundary and executed as same experiments as described in Section 3.A. As shown in Fig. 4(c), with variations of lesion segmentation accuracy (Dice score (Xian *et al.*, 2018)) between real-boundary and fake-boundary from around 95% to 70%, the classification results of BIRADS-SSDL method appear slight drop before Dice is greater than 90%, and then there are significant declines. This indicated that BIRADS-SSDL is robust on slight deviation of the lesion's boundary. It does not have too strict requirements for tumor boundary, and potential alleviate the difficult problem of lesion segmentation.

### 3.5 Effects of Gaussian Filter Parameter σ

Fig. 5 shows the variations in the overall accuracy of classification results for the two BIRADS-based methods across σ values. It can be seen that BIRADS-SSDL has higher accuracy than BIRADS-SCAE across almost every σ value, though it has some fluctuations. All curves show the best result when σ = 20 and decrease slightly with smaller or larger σ values. It also can be seen that the standard deviation variations are relatively small, around σ = 20. This indicates that the area across the lesion boundary within a certain range plays an important role in diagnosis and should be given more attention.

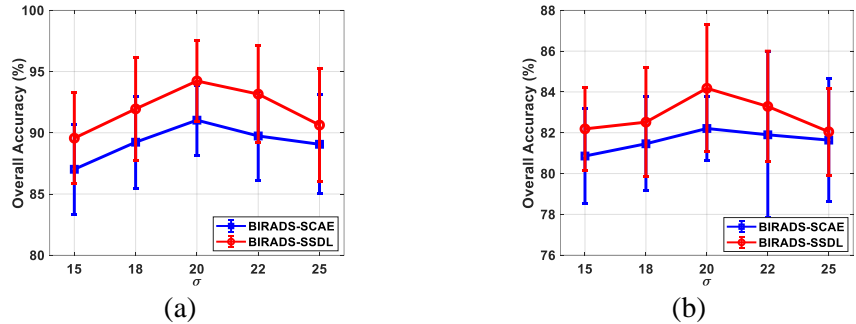

Figure 5. Effect of parameter σ in a Gaussian filter on BIRADS-based methods for (a) UDIAT and (b) UTSW dataset.

Table 6. The performance summary of breast ultrasound CAD system

| Ref. | Dataset (benign/ malignant) | Availability | Features / Classifier | Performance (%) | | |
|---|---|---|---|---|---|---|
| | | | | ACC | SEN | SPE |
| (Singh *et al.*, 2016) | 88 / 90 | Private | Manual feature / BPNN | 95.86 | 95.14 | 96.58 |
| (Han *et al.*, 2017) | 4254/3154 | Private | GoogLeNet | 91.23 | 84.29 | 96.07 |
| (Zhang *et al.*, 2016) | 135 / 92 | Private | Boltzmann machine | 93.40 | 88.60 | 97.10 |
| (Cheng *et al.*, 2016) | 275 / 245 | Private | SDAE | 82.40 | 78.70 | 85.70 |
| (Shi *et al.*, 2016) | 100 / 100 | Private | DPN / SVM | 92.40 | 92.67 | 91.36 |
| (Antropova *et al.*, 2017) | 1978 / 415 | Private | Manual feature /VGG 19 | AUC = 90.00 | | |
| (Prabusankarlal *et al.*, 2015) | 70 / 50 | Public (GVHE) | Manual feature / SVM | 95.83 | 96.00 | 95.71 |
| (Byra, 2018) | 48 / 52 | Public (OASBUD) | VGG 19 / SVM | 80.00 | 80.80 | 79.20 |
| (Byra *et al.*, 2018) | 110 / 53 | Public (UDIAT) | VGG 19 / FLDA | 84.00 | 85.51 | 83.40 |
| **Our** | **83 / 45** | **Public (UDIAT)** | **BFM / SSDL** | **94.23** | **82.99** | **98.61** |
| | **178/ 80** | **Private** | **BFM / SSDL** | **84.38** | **60.83** | **93.82** |

## 4 Discussion and Conclusions

We developed a novel BIRADS-SSDL network to incorporate clinically-assigned breast lesion characteristics into a task-oriented semi-supervised deep learning method for accurate diagnosis on US images with a relatively small training dataset. We verified the effectiveness of BIRADS-SSDL on two breast US image datasets and found that the network achieved high diagnostic accuracy. In the public UDIAT dataset, the BIRADS-SSDL network achieved the highest ACC and AUC values of 94% and 90%, respectively. In the in-house clinical dataset, we obtained ACC and AUC values of 84% and 77%, respectively.

Here, we provide a comparison with state-of-the-art CAD methods in Table 6. Unlike those traditional machine learning methods (Singh *et al.*, 2016; Prabusankarlal *et al.*, 2015; Jalalian *et al.*, 2013), the proposed BIRADS-SSDL method automatically learned representative and discriminative features by hierarchical deep neural network. Different from the recent DL methods with pre-training techniques or transfer learning (Antropova *et al.*, 2017; Byra, 2018; Byra *et al.*, 2018), we fuse the existing conventional BIRADS features into a semi-supervised deep neural network which improves performance in breast lesion diagnosis. In our experiments, BIRADS-SSDL used 102 labeled images for training the network and achieved an AUC (~90%) comparable to the results reported in recent papers (Antropova *et al.*, 2017; Byra, 2018), where the AUC values are around 85% and the networks are trained on bigger datasets. On the same public dataset (UDIAT, discard 35 images with size smaller than the default input size of the network), the ACC value of BIRADS-SSDL was higher than the highest ACC value of the previous transfer DL method (Byra *et al.*, 2018), 94% vs. 84%. Although it is hard to fairly evaluate the performance of different methods utilizing different datasets, our method is promising method for effective breast ultrasound CAD.

We evaluated the generalizability of BIRADS-SSDL with experiments across two datasets collected from two different institution/and US devices. When training the model on both datasets together, the developed BIRADS-SSDL is generalizable across the different US devices and institutions without overfitting to a single dataset and achieved satisfactory results. Furthermore, we investigate the effects of two key strategies on the performance of the proposed method, including lesion contour accuracy and Gaussian filter parameter. Experimental results showed that pre-training strategy can help the proposed speed up convergence without improving classification accuracy. And the experiments of the boundary accuracy proved the proposed method could achieve a satisfactory performance when there are slight deviation (Dice score ≥ 90%) of the lesion's boundary.

There were several limitations to our study. According to the American College of Radiology and clinical experience, radiologists usually characterize tumors with five-point, including shape, orientation, margin, echo pattern, and posterior features. Our semi-supervised deep learning model is guided by BFMs for automatically extracting representative features which potentially emphasized pathological results high-correlated features, such as shape and margin features. Though orientation, echo pattern, and posterior features were still embedded in BFMs, they are not explicitly enhanced. Our future work will incorporate new image processing methods to enhance orientation, echo pattern and posterior acoustic features in the US image and integrate the processed images into deep learning for further improve diagnosis accuracy. Our method does not take into account the relationship between lesion images with multiple different angles from the same patient. In the future, we will develop an end-to-end semi-supervised breast US diagnosis ensemble system that includes lesion segmentation and classification, which will not only fuse the clinical lesion characteristics but also use multiple US images from the same patient to make a joint decision. Furthermore, it can be seen that the datasets used by state-of-the-art studies are different and have huge differences in the size and the modality. Most of datasets used in the studies mentioned above is still small and private. In the future, we will verify our method on the bigger dataset and share our in-house dataset to help the research in this area.

   In summary, the proposed BIRADS-SSDL achieves the best results among the compared methods in each case and has the capacity to deal with multiple different datasets under one model, thereby indicating that BIRADS-SSDL is a promising method for effective breast US lesion CAD using small datasets.